\documentclass[aps,pra,twocolumn,floatfix,showpacs,amsmath,amssymb,superscriptaddress]{revtex4-1}

\usepackage{amsthm}
\usepackage{graphicx}
\usepackage{xcolor}
\usepackage{epstopdf}
\usepackage{subfigure}
\usepackage{capt-of}
\usepackage{amsmath,amssymb}
\usepackage[colorlinks=true,citecolor=blue,urlcolor=blue]{hyperref}
\usepackage{bm}
\usepackage{mathtools}
\makeatletter

\newcommand{\Rmnum}[1]{\expandafter\@slowromancap\romannumeral #1@}
\makeatother
\newcommand{\be}{\begin{equation}}
\newcommand{\ee}{\end{equation}}
\newcommand{\ba}{\begin{eqnarray}}
\newcommand{\ea}{\end{eqnarray}}
\newcommand{\ban}{\begin{eqnarray*}}
\newcommand{\ean}{\end{eqnarray*}}
\newcommand{\one}{\leavevmode\hbox{\small1\normalsize\kern-.33em1}}

\begin{document}
\title{One-dimensional Polar Spinor Droplets}

\author{Hao Zhu}
\affiliation{Liberal Study College, Beijing Institute of Petrochemical Technology, Beijing, China}
\author{Wen-Kai Bai}
\affiliation{Shaanxi Key Laboratory for Theoretical Physics Frontiers, Institute of Modern Physics, Northwest University, Xi'an 710127, China}
\author{Yi-Ran Shen}
\affiliation{Liberal Study College, Beijing Institute of Petrochemical Technology, Beijing, China}
\author{Xiao-Fei Zhang}
\email{xfzhang@sust.edu.cn}
\affiliation{School of Physics and Information Science, Shaanxi University of Science and Technology, Xi'an 710021, China}
\author{Wu-Ming Liu}
\email{wmliu@iphy.ac.cn}
\affiliation{Tsientang Institute for Advanced Study, Hangzhou 310024, Zhejiang, China}
\affiliation{Beijing National Laboratory for Condensed Matter Physics and Institute of Physics, Chinese Academy of Sciences, Beijing 100190, China}
\author{Boris A. Malomed}
\email{malomed@tauex.tau.ac.il}
\affiliation{Department of Physical Electronics, School of Electrical Engineering, Faculty of Engineering, Tel Aviv University, Tel Aviv 69978, Israel}
\affiliation{Instituto de Alta Investigaci\'on, Universidad de Tarapac\'a, Casilla 7D, Arica, Chile}

\begin{abstract}
We derive a channel-resolved Lee-Huang-Yang correction and construct an extended Gross-Pitaevskii model for one-dimensional polar spin-1 quantum droplets. The fluctuation contribution separates into density and spin channels, which supports self-bound droplets even when the spin-independent mean-field interaction is repulsive. Stationary solutions exhibit a continuous crossover from soliton-like to flat-top droplets, accompanied by saturation of the chemical potential and peak density as the particle number increases. Within the parameter range examined here, linear Bogoliubov analysis together with weak-perturbation dynamics supports the stability of both droplet types. A quadratic-Zeeman quench reveals a finite-size crossover in breathing dynamics and distinct nonequilibrium roles of the density and spin fluctuation channels. Representative head-on collisions further show that the finite-size crossover modulates phase-sensitive nonlinear scattering, with in-phase impact producing coalescence-like central retention and out-of-phase impact favoring quasi-elastic separation. The analysis clarifies how density and spin fluctuations shape equilibrium structure and nonequilibrium response in low-dimensional polar spinor droplets.
\end{abstract}
%\pacs{}

\date{\today}
\maketitle
\section{Introduction}
Quantum droplets (QDs) are ultradilute self-bound quantum fluids in which a residual mean-field attraction is balanced by the repulsive Lee-Huang-Yang (LHY) correction from quantum fluctuations~\cite{LHY1957}. Petrov provided the key theoretical prediction that Bose-Bose mixtures near mean-field collapse can be
stabilized into liquidlike droplets by the LHY term~\cite{Petrov2015}. Experiments on strongly dipolar dysprosium gases have produced the first examples of fluctuation-stabilized QDs under the action of competing contact and dipole-dipole interactions~\cite{Ferrier2016,Schmitt2016}. The same stabilization mechanism was later confirmed in potassium gases
with purely contact interactions, thereby verifying Petrov's prediction~\cite{Cabrera2018,Semeghini2018}. These experiments established the formation of QDs in dilute gases as an accessible beyond-mean-field phenomenon and opened the way to quantitative studies of collective and nonequilibrium dynamics, including LHY-dominated collective motion in binary QDs~\cite{Skov2021,He2023,Cavicchioli2025PRL} and diagnostics based on excitation spectra and time-dependent evolution~\cite{Nilsson2022,Englezos2023,Gangwar2024,Zhu2024RydbergPRR}.

Compared with scalar and binary-mixture systems, multicomponent spinor condensates offer extra interaction channels generated by spin fluctuations and internal symmetries~\cite{StamperKurn2013,Kawaguchi2012}. At the mean-field level, spinor condensates also support polar and ferromagnetic solitons, spin textures, and Zeeman-controlled defects~\cite{Song2013Spin1Review,Li2005SpinorSoliton,Zhao2015ZeemanDefects}, which provide useful reference points for distinguishing fluctuation-stabilized droplets from conventional spinor nonlinear waves. For polar spin-1 condensates, these fluctuations can support self-trapped states beyond the mean-field collapse,
and the quadratic Zeeman shift determines both the QD-formation threshold and the equilibrium density~\cite{Yogurt2022Spin1}. The importance of internal control parameters extends beyond this polar case, as polarized spinor and Rabi-coupled mixtures also show that Zeeman and coupling parameters modify the QDs' stability and phase boundaries~\cite{Yogurt2023Polarized}. More broadly, three-component systems display genuine multibody binding, including many-body bound states facilitated by Borromean binding, and can host core-shell self-bound configurations in realistic mixtures~\cite{Ma2021Borromean,Ma2025Shell}. Even with these advantages, many three-dimensional spinor-QD schemes still rely on negative spin-independent density interaction, $c_0<0$, together with the delicate interchannel balance, which makes the required conditions more restrictive in experiments~\cite{LiSaito2024PRR}.

One-dimensional (1D) QDs are of particular interest because quantum fluctuations are enhanced and density scaling differs from that in 3D, leading to stabilization mechanisms that are absent in the standard 3D droplet framework~\cite{CuiMa2021Confinement,Edler2017,Roccuzzo2019Tube,Ilg2023Dipolar}. In confined 1D geometries, the beyond-mean-field contribution can dominate in the low-density regime and support ultradilute self-bound states. In this setting, the distinction between soliton-like bound states and superfluid QDs becomes a central issue because finite-size effects and beyond-mean-field attraction can favor different configurations~\cite{Du2023,Tylutki2020}. Excitation spectra then provide quantitative information on compressibility and linear stability~\cite{Nilsson2022,Englezos2023}, while quench-induced breathing oscillations probe finite-size response and nonequilibrium compressional dynamics~\cite{Gangwar2022,Orignac2024Breathing}. However, for polar spin-1 QDs in this regime, a unified theory connecting channel-resolved LHY corrections with stability and quench dynamics is still lacking~\cite{Englezos2024Imbalanced}.

In this work, we derive a channel-resolved LHY correction for 1D polar spin-1 QDs and formulate the corresponding extended Gross-Pitaevskii (eGPE) within a self-consistent quasi-1D parameter range defined below.  The present treatment is restricted to the polar manifold. A full spinor extension would require coupled evolution of \((\psi_{+1},\psi_0,\psi_{-1})^T\) and an LHY functional for the local spin composition, which is outside the scope of this work. The resulting description separates density and spin fluctuation channels, treats the quadratic Zeeman term as a direct control parameter, and supports self-binding even when the 1D spin-independent mean-field interaction is repulsive. We then examine equilibrium morphology, linear stability, quench-induced breathing dynamics, and head-on collisions with controlled relative phase. The collision dynamics tests how the finite-size crossover redistributes injected translational kinetic energy. This formulation also distinguishes how density and spin fluctuations affect both self-binding and driven dynamics in polar spin-1 droplets~\cite{GuCui2023,Nikolaou2024PRA}.

\section{EGPE for polar phase spin-1 BECs}
The derivation of the polar-phase channel-resolved LHY correction and the quasi-1D reduction is given in Appendix~\ref{app:lhy_derivation}. Here we retain only the effective dimensionless 1D eGPE used in the numerical calculations.

Within the local-density approximation, the homogeneous equation of state is evaluated at the local dimensional 1D density \(n_{\rm 1D}(x)=|\psi(x,t)|^2\), and the eGPE follows from \(\delta E/\delta\psi^*\). The local LHY chemical potential \(\mu_{\rm LHY}^{1D,P}(n_{\rm 1D})\) contains a density-fluctuation contribution and a spin-fluctuation contribution governed by the auxiliary function \(S_{1D}\) defined in Appendix~\ref{app:lhy_derivation}.
Using a reference density $n_1$, we define the characteristic length scale $\xi_1$, energy scale $E_1$, and time scale $t_0=\hbar/E_1$. Following Ref.~\cite{Yogurt2022Spin1}, we define $n_1$ as the $q\to0$ limit of the equilibrium density $n_0(q)$, which is obtained from the vanishing-pressure
condition of the homogeneous energy density. In practice, $n_1$ is obtained numerically as the $q\to0$ root of \(P(n_{\rm 1D})=n_{\rm 1D}\mu(n_{\rm 1D})-\mathcal E(n_{\rm 1D})=0\) for the same equation of state as used below. It serves only as the normalization density. With $E_1=|c_0^{1D}|\,n_1/3$, $t_0=\hbar/E_1$, and $\xi_1=\sqrt{3\hbar^2/(M|c_0^{1D}|\,n_1)}$, we define
$\psi=\sqrt{n_1}\Phi(\tilde x,\tau)$ with $\tilde x=x/\xi_1$ and $\tau=t/t_0$.  The dimensionless norm is \(\tilde N=\int d\tilde x\,|\Phi(\tilde x)|^2\), whereas the corresponding physical atom number is \(N_{\rm phys}=n_1\xi_1\tilde N\). Hereafter, tildes are omitted, \(\tau\) is written as \(t\), and the dimensionless norm is denoted by \(N\). Thus, the values of \(N\) used below should not be read as literal atom counts.
After division by the energy scale \(E_1\), the LHY chemical-potential correction obtained from Eq.~\eqref{eq:lhy_mu_1d_polar} takes the \(n_1\)-normalized dimensionless form \(\mu_{\rm LHY,1}^{1D,P}(n)=-\eta_0\sqrt{n} +\eta_1\sqrt{n}\,S_{1D}\!\left(\tau_q/n\right)\), where \(n=|\Phi|^2\) is now the dimensionless density. The subscript \(1\) specifies normalization with respect to the reference density \(n_1\) rather than a separate LHY contribution. The dimensionless parameters are \(\tau_q\equiv q/(n_1 c_1^{1D})\), \(\gamma=\tau_q/n\), \(\eta_0=(3/\pi)(\sqrt{M}/\hbar)\sqrt{|c_0^{1D}|/n_1}\), and
$\eta_1=(3/\pi)(\sqrt{M}/\hbar)(c_1^{1D})^{3/2}/(|c_0^{1D}|\sqrt{n_1})$. The resulting dimensionless eGPE reads
\begin{equation}
i\,\partial_t\Phi=
\left[
-\frac12\partial_x^2
+3\,|\Phi|^2
+\mu_{\rm LHY,1}^{1D,P}\!\big(|\Phi|^2\big)
\right]\Phi.
\label{eq:egpe_dimless}
\end{equation}
In Eq.~\eqref{eq:egpe_dimless}, \(\Phi(x,t)\) is the dimensionless condensate order parameter,
\(x\) and \(t\) are dimensionless coordinate and time, and
\(n\equiv|\Phi|^2\) is the local dimensionless density. The term
$3|\Phi|^2$ is the normalized mean-field contribution, and
$\mu_{\rm LHY,1}^{1D,P}(n)$ is obtained by rescaling the dimensional LHY chemical-potential correction in Eq.~\eqref{eq:lhy_mu_1d_polar}.
Numerical solutions for the QD states and their stability and dynamical properties are presented below.
\section{Numerical solutions for 1D QDs }
\subsection*{A. Soliton-like QDs and flat-top QDs}
\begin{figure}[t]
\centering
\includegraphics[width=\linewidth]{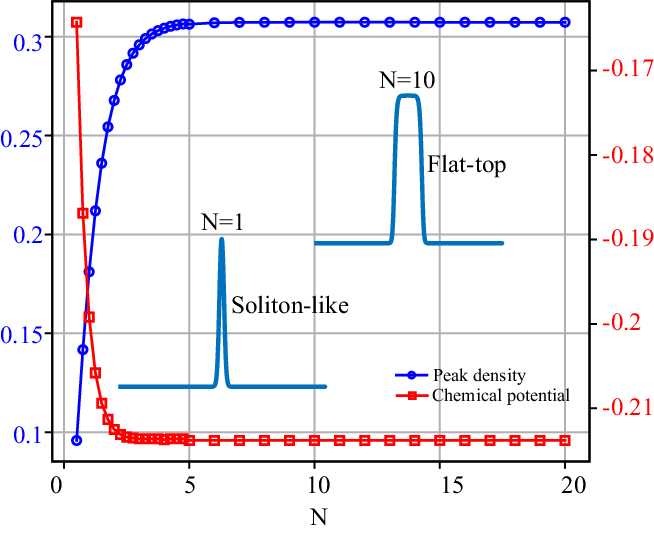}
\caption{Dimensionless chemical potential $\mu$ and peak density $\rho_{\rm max}$ as functions of particle number $N$ for 1D polar spin-1 QDs. The inset profiles correspond to the soliton-like and flat-top profiles at $N=1$ and $N=10$, respectively. For the morphological classification used here, $f_{\rm flat}=\sigma(\rho_{\rm c})/\langle\rho_{\rm c}\rangle$ quantifies the central flatness from the central density values $\rho_{\rm c}$, evaluated over $\rho\ge0.9\,\rho_{\rm max}$, and $f_{\rm edge}=\langle\rho_{\rm edge}\rangle/\rho_{\rm max}$ measures the relative edge density. These indicators are used only for operational labeling along the finite-size crossover. The flat-top class is defined by $f_{\rm flat}<0.05$ and $f_{\rm edge}<10^{-3}$, while the remaining plotted states are labeled soliton-like.}
\label{fig:fig1}
\end{figure}

Stationary solutions of the dimensionless eGPE [Eq.~\eqref{eq:egpe_dimless}] are computed by means of imaginary-time propagation algorithm ($t\to-it$) implemented with the help of the split-step Fourier method~\cite{Du2023,Tylutki2020}. The local LHY term is evaluated with the approximation $S_{1D}(\gamma)\approx-2/\sqrt{1+(8/\pi^2)\gamma}$, where $\gamma=\tau_q/n$ and $n=|\Phi|^2$ is the dimensionless local density.

 Figure~\ref{fig:fig1} summarizes the finite-size evolution of the stationary branch through the chemical potential, peak density, and representative density profiles. The stationary self-bound solutions are characterized by the chemical potential $\mu=\langle\Phi|\hat H_{\rm eff}|\Phi\rangle/N$ and the peak density $\rho_{\rm max}=\max|\Phi|^2$, where $\hat H_{\rm eff}$ is the effective eGPE Hamiltonian in Eq.~\eqref{eq:egpe_dimless}. The dimensionless chemical potential $\mu(N)$ decreases monotonically, remains negative throughout the calculated range, and approaches $\mu\simeq-0.21$ for $N\gtrsim3$. This behavior is consistent with self-binding, for which the equilibrium with vacuum occurs at negative chemical potential~\cite{Petrov2015,Cabrera2018,Semeghini2018}. At the same time, $\rho_{\rm max}(N)$ approaches $\rho_{\rm max}\simeq0.31$ for $N\gtrsim4$, indicating the emergence of a well-defined bulk density~\cite{Englezos2023,Du2023}. These features indicate that the finite-size solutions approach the bulk zero-pressure state defined by the equation of state in Sec.~II, so that increasing $N$ enlarges the droplet primarily through spatial growth rather than by a further increase of the central density. The approach to the equilibrium bulk density is governed by the competition between the repulsive mean-field pressure and attractive beyond-mean-field fluctuation-induced pressure towards the zero-pressure condition of the equation of state.

The corresponding density profiles evolve continuously from soliton-like states at small $N$, where the surface-to-bulk ratio is large, to flat-top QDs at larger $N$, where an extended core is bounded by narrow edge regions. The representative profiles in Fig.~\ref{fig:fig1}, which correspond to $N=1$ and $N=10$, are consistent with established QD behavior in low-dimensional and mixture settings~\cite{Tarruell2018,Kartashov2019,Caldara2022,Sturmer2021}. The monotonic decrease of $\mu(N)$ corresponds to $d\mu/dN<0$, or equivalently $dN/d\mu<0$ along the computed branch, and therefore satisfies the Vakhitov--Kolokolov necessary stability criterion for bright self-bound states~\cite{Vakhitov1973}. The simultaneous saturation of $\rho_{\rm max}(N)$ is consistent with previous results for 1D droplets~\cite{Englezos2023,Du2023}. In Fig.~\ref{fig:fig1}, soliton-like and flat-top are operational labels assigned by the central-flatness and edge-density criteria in the caption.  They mark a finite-size crossover toward a bulk-like profile, not a sharp critical particle number or a separate phase boundary. The saturation of \(\mu\) and \(\rho_{\rm max}\) supports this crossover interpretation. In the polar spin-1 setting, this equilibrium sequence shows that self-binding remains available even when the spin-independent mean-field interaction is repulsive, thereby relaxing a key constraint of many three-dimensional spinor-droplet scenarios~\cite{Yogurt2022Spin1,Yogurt2023Polarized}.

\subsection*{B. Stability diagnostics of 1D QDs}
Dynamical stability is assessed by linear Bogoliubov-de Gennes (BdG) spectroscopy around the stationary state and by real-time evolution in the presence of a weak perturbation, which are two complementary diagnostics widely used in the studies of QDs~\cite{Englezos2023,Tylutki2020}. The BdG spectrum identifies unstable channels at the linear level, whereas the real-time evolution tests whether weak perturbations remain bounded over long times. For self-bound droplets, the combined stability criterion is required because low-lying discrete modes can approach the continuum threshold for particle emission.

For the linear stability analysis, the condensate field is expanded around the stationary solution as $\Phi(x,t)=\big[\phi(x)+\lambda\big(u(x)e^{-i\omega t}+v^*(x)e^{i\omega t}\big)\big]e^{-i\mu t}$ with $0<\lambda\ll1$, where $\phi(x)$ is the stationary state, $u(x)$ and $v(x)$ denote the BdG mode amplitudes, while $\omega$ is the eigenfrequency. Substitution of this ansatz into Eq.~\eqref{eq:egpe_dimless}, retaining only terms up to first order in the perturbation amplitude, yields the coupled BdG equations
\begin{equation}
\omega u=\hat L u+\hat M v,\qquad
-\omega v=\hat L^* v+\hat M^* u,
\end{equation}
where $\hat L=-\frac12\partial_x^2-\mu+G(|\phi|^2)+|\phi|^2G'(|\phi|^2)$, $\hat M=G'(|\phi|^2)\phi^2$, $G(n)=3n+\mu_{\rm LHY,1}^{1D,P}(n)$, and \(G'(n)=dG/dn\). For the real stationary profiles and real differential operator used below, \(\hat L^*=\hat L\) and \(\hat M^*=\hat M\). Eigenmodes with $\operatorname{Im}(\omega)>0$ indicate dynamical instability, whereas a purely real BdG spectrum indicates linear stability~\cite{Nilsson2022,Du2023}. The linearized equations are then solved as a BdG matrix eigenvalue problem.

The real-time evolution of weakly perturbed stationary states is then used to assess the linear stability prediction by monitoring spatial localization and bounded oscillatory motion. Agreement between the BdG spectrum and direct dynamics supports the stability diagnosis near spectral thresholds~\cite{Englezos2023}. In Fig.~2, the density- and spin-channel LHY coefficients are fixed as $(\eta_0,\eta_1)=(1,1)$, the dimensionless quadratic-Zeeman parameter is fixed at $\tau_q=1$, and only the particle number is varied through $N=1,5,10$, therefore the changes across the figure are attributed to finite-size effects.  Apart from the expected near-zero phase and translational modes, the nonzero-frequency BdG eigenvalues shown in Fig.~2 have no resolvable imaginary part. Together with the bounded weak-perturbation dynamics, this supports linear stability in the explored regime.

\begin{figure}[t]
\centering
\includegraphics[width=\linewidth]{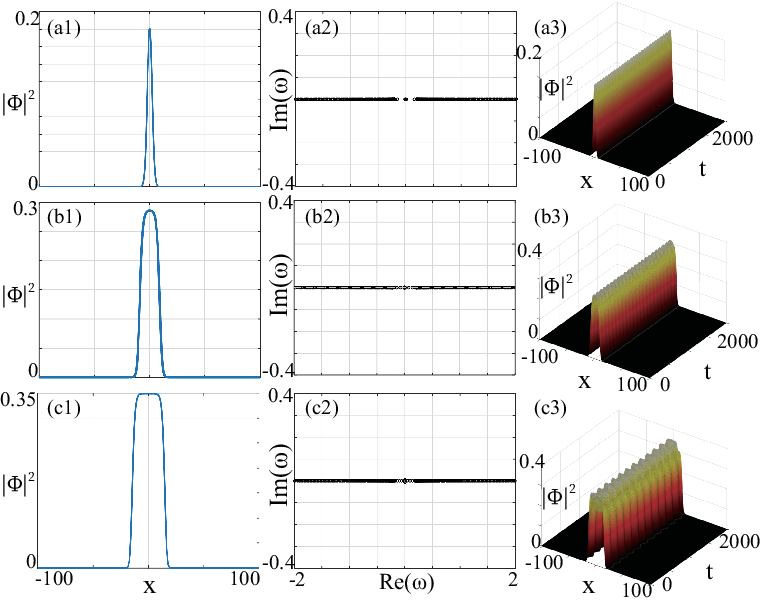}
\caption{Stationary densities $|\Phi(x)|^2$, BdG spectra, and the weakly perturbed real-time evolution of 1D polar spin-1 QDs at fixed dimensionless density- and spin-channel LHY coefficients $(\eta_0,\eta_1)=(1,1)$ and fixed dimensionless quadratic-Zeeman parameter $\tau_q=1$. The three rows correspond to particle numbers $N=1$, $5$, and $10$. In each row, the left, middle, and right panels show $|\Phi(x)|^2$, the BdG spectrum, and the real-time evolution after a weak random perturbation of amplitude $\epsilon=0.01$.}
\label{fig:fig2}
\end{figure}

Perturbation-amplitude tests at $\epsilon=0.005$, $0.01$, and $0.02$ give the same stability classification under weak perturbations. Because the channel parameters remain fixed at $(\eta_0,\eta_1,\tau_q)=(1,1,1)$, the changes across Fig.~2 are driven only by the particle number. The case of $N=1$ [panels (a1--a3)] shows a sparse set of low-lying discrete modes for the soliton-like state. The case of $N=5$ [panels (b1--b3)] contains a denser low-lying sector and exhibits the strongest envelope modulation in the real-time evolution. The case of $N=10$ [panels (c1--c3)] corresponds to a flat-top QD and exhibits the most regular long-time oscillatory response.  In all three cases, the same conclusion holds for the nonzero-frequency BdG eigenvalues shown in Fig.~2. This particle-number dependence agrees with the crossover from soliton-like to flat-top profiles identified in Fig.~1.  Appendix~\ref{app:bdgN_spectrum} gives the \(N\)-dependent finite-frequency BdG branches and a semi-analytic check. The appendix provides supporting spectral evidence for the stability and breathing analysis in the main text. In particular, the lowest even-parity branch remains below the particle-emission threshold and follows the expected large-\(N\) compressional scale. In the polar spin-1 setting with $c_0^{1D}>0$, the explored parameter range supports stable QDs across the crossover from soliton-like to flat-top profiles and leaves channel tuning available for probing a nonequilibrium response~\cite{Gangwar2022,Orignac2024Breathing,Du2023}.
\subsection*{C. Quench-induced breathing dynamics}
To examine the nonequilibrium response in the linearly stable regime identified above, we induce breathing dynamics by a sudden quench of the dimensionless quadratic-Zeeman parameter $\tau_q$ at $t=0$. Each trajectory in Fig.~3 starts from the stationary state at the prequench value $\tau_q=1$ and evolves after $\tau_q$ was dropped to $0$. At fixed particle number, this protocol excites collective breathing motion and probes the compressional response of the self-bound state. The response is characterized by the normalized second central moment of the density profile, $\sigma_x^2(t)/\sigma_x^2(0)$, with $\sigma_x^2(t)\equiv\langle x^2(t)\rangle-\langle x(t)\rangle^2$~\cite{Mistakidis2021,Tanzi2019}. Because this observable is normalized by $\sigma_x^2(0)$ and removes center-of-mass drift by construction, it directly tracks the breathing-induced change of the droplet width across different particle numbers and channel parameters. The following discussion concerns a breathing-dominated collective response rather than the identification of a single nonlinear mode.

The initial states and the quench protocol are parity symmetric, so the displayed evolution preserves $\langle x(t)\rangle=0$ and $\langle x^2(t)\rangle/\langle x^2(0)\rangle=\sigma_x^2(t)/\sigma_x^2(0)$. Two single-parameter scans are carried out at fixed $N=1,5,10$. Panels (a1--a3) vary the dimensionless spin-channel LHY coefficient $\eta_1=1,2,3$ at fixed density-channel coefficient $\eta_0=1$, whereas panels (b1--b3) vary the dimensionless density-channel LHY coefficient $\eta_0=1,2,3$ at fixed spin-channel coefficient $\eta_1=1$. Unlike earlier QD-quench studies, which mainly varied global interaction or spin-orbit-coupling parameters in scalar or mixture settings~\cite{Gangwar2024,Gangwar2022,Mistakidis2021}, the present 1D polar spin-1 formulation contains two independent channel control knobs and therefore distinguishes the nonequilibrium roles of density and spin fluctuations.

\begin{figure}[t!]
\centering
\includegraphics[width=0.92\columnwidth]{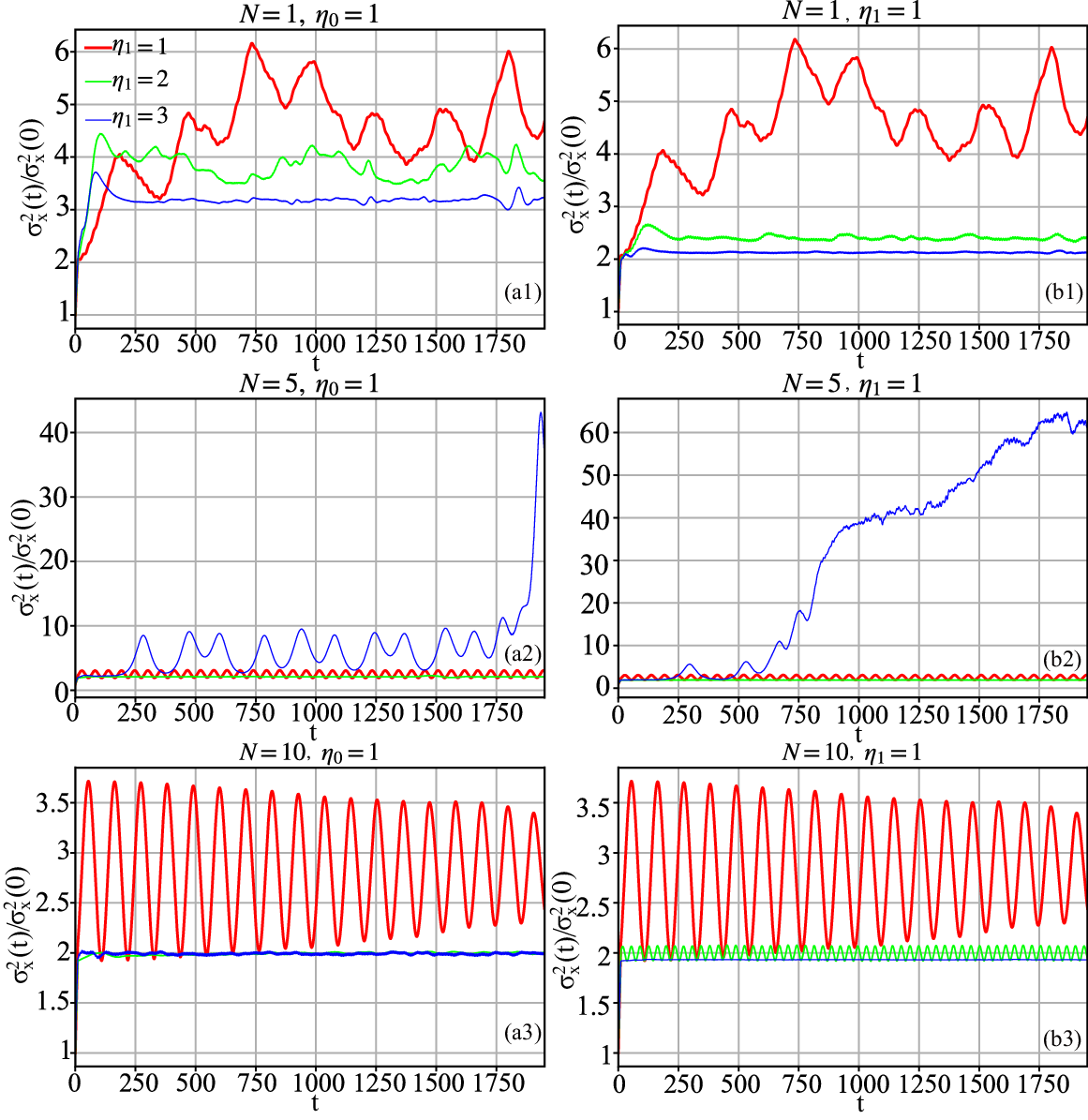}
\caption{The evolution of the normalized density-width variance $\sigma_x^2(t)/\sigma_x^2(0)$, which measures the QD's width relative to its initial value, after a sudden quench of the dimensionless quadratic-Zeeman parameter $\tau_q$ at $t=0$. For the parity-symmetric trajectories shown here, this ratio equals \(\langle x^2(t)\rangle/\langle x^2(0)\rangle\). The initial state in each case is the stationary solution at $\tau_q=1$, and the subsequent dynamics is generated after the quench to $\tau_q=0$. Panels (a1--a3) are obtained at fixed dimensionless density-channel LHY coefficient $\eta_0=1$ with spin-channel LHY coefficient $\eta_1=1,2,3$, whereas panels (b1--b3) are obtained at fixed $\eta_1=1$ with $\eta_0=1,2,3$. The three rows correspond to particle numbers $N=1$, $5$, and $10$.}
\label{fig:fig3}
\end{figure}

The breathing response at $N=1$, $5$, and $10$ exhibits a finite-size crossover governed by the competition between surface and bulk restoring mechanisms~\cite{Du2023,Tylutki2020}. For $N=1$ [panels (a1) and (b1)], the soliton-like state exhibits large but bounded oscillations whose amplitude varies clearly with the channel coupling, which indicates that the restoring response at small $N$ remains dominated by interfacial effects. For $N=5$ [panels (a2) and (b2)], where interfacial and bulk-compressional contributions are comparable, the response exhibits the strongest envelope modulation and the most pronounced long-time enhancement at the largest couplings. Combined with the denser low-lying sector in Fig.~2  and the \(N\)-dependent spectrum in Appendix~\ref{app:bdgN_spectrum}, this behavior is compatible with stronger coupling of the breathing-dominated response to nearby low-lying compressional modes in this intermediate-size case. For $N=10$ [panels (a3) and (b3)], corresponding to the flat-top state, the oscillations are more regular and the envelope is tighter under the same parameter variation, which indicates a more bulk-compressional restoring response. This spectral comparison serves only as supporting evidence for the observed envelope modulation and regularity, rather than as a direct identification of the nonlinear modes participating in the quench dynamics~\cite{Englezos2023,Gangwar2022,Orignac2024Breathing}. This particle-number dependence agrees with the crossover from soliton-like to flat-top profiles identified in Fig.~1.

At fixed $N$, we compare the quench response under variation of the density-channel LHY coefficient $\eta_0$ [panels (b1--b3)] and the spin-channel LHY coefficient $\eta_1$ [panels (a1--a3)]. In the explored range, $\mu_{\rm LHY,1}^{1D,P}(n)=-\eta_0\sqrt n+\eta_1\sqrt n\,S_{1D}(\tau_q/n)$ with $S_{1D}(\tau_q/n)<0$, so raising either coefficient increases the magnitude of the attractive beyond-mean-field contribution and modifies the effective compressional response probed by the quench dynamics~\cite{Englezos2023,Yogurt2023Polarized,Lavoine2021}. Variation of $\eta_0$ mainly shifts the background compressional response, whereas variation of $\eta_1$ more strongly affects envelope modulation. The distinction is drawn from the channel-dependent trajectories rather than from an explicit decomposition into nonlinear collective modes. Over the displayed evolution time, the responses for $N=1$ and $N=10$ remain bounded and localized, whereas the $N=5$ case exhibits the strongest long-time amplification without resolvable dynamical instability. That behavior is compatible with the crossover regime, where interfacial and bulk restoring mechanisms are both important and the denser low-lying sector favors stronger beating between nearby collective modes. The density-channel coefficient $\eta_0$ sets the compressional background scale, whereas the spin-channel coefficient $\eta_1$ more strongly affects envelope modulation in the intermediate-size regime~\cite{Englezos2023,Gangwar2022,Orignac2024Breathing}.

\subsection*{D. Head-on collision dynamics}
 We finally use head-on collisions to test how relative phase and droplet morphology shape the post-collision density pattern. This collision setup follows standard one-dimensional and quasi-one-dimensional studies based on two stationary droplets with opposite momenta and a controlled relative phase~\cite{Astrakharchik2018Droplets,Otajonov2024Quasi1D,AlKhawaja2024FlatTop}. The initial state is constructed as
\begin{equation}
\Phi(x,0)=A_N\left[
\phi_N(x+x_0)e^{iv_0x}+e^{i\delta}\phi_N(x-x_0)e^{-iv_0x}
\right].
\end{equation}
Here $\phi_N$ is the stationary solution of Eq.~\eqref{eq:egpe_dimless} with norm $N$, $x_0$ is the initial half-separation between the two droplets, and $A_N$ is chosen so that $\int_{-\infty}^{\infty} dx\,|\Phi(x,0)|^2=2N$. Parameters $v_0$ and $\delta$ are the dimensionless incidence speed and relative phase, respectively. The factors $e^{\pm i v_0 x}$ impart opposite momenta to the two droplets. The channel parameters are fixed as $(\eta_0,\eta_1,\tau_q)=(1,1,1)$, with $\delta=0$ and $\delta=\pi$ specifying the in-phase and out-of-phase head-on geometries, respectively.  The representative physical scale of these dimensionless collision parameters follows from the mapping given in Appendix~\ref{app:lhy_derivation}.

The post-collision classifications in Table~\ref{tab:collision_diagnostics} are based on density diagnostics evaluated at the final displayed time $t_f$. The central-retention fraction is $f_c=N_{\rm cen}(t_f)/(2N)$, with $N_{\rm cen}(t_f)=\int_{|x|<\max(8,0.35x_0)} dx\,|\Phi(x,t_f)|^2$. The quantity $f_r$ denotes the delocalized radiative fraction after excluding the central remnant and the dominant outgoing packets. These operational diagnostics are consistent with post-collision analyses that quantify central remnant population, outgoing-peak separation, and radiated norm~\cite{Ferioli2019Collisions,Cikojevic2021Collisions,Hu2022SymmetricCollisions,AlKhawaja2024FlatTop}. Selected velocity, phase, box-size, and grid-resolution runs were used to check the robustness of these classifications.

\begin{figure}[t]
\centering
\includegraphics[width=\columnwidth]{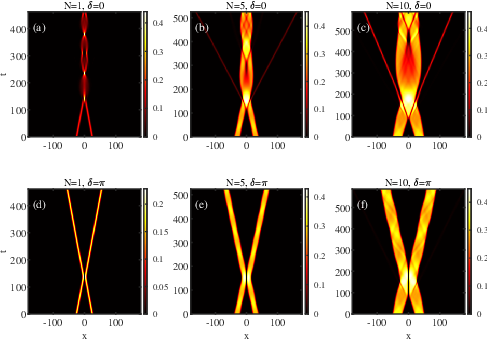}
\caption{The density evolution in head-on collisions between two identical stable 1D polar spin-1 QDs at $v_0=0.16$ and $(\eta_0,\eta_1,\tau_q)=(1,1,1)$. Panels (a)--(c) show $\delta=0$ with $N=1$, $5$, and $10$. Panels (d)--(f) show $\delta=\pi$ for the same particle numbers.}
\label{fig:fig4}
\end{figure}

\begin{table}[t]
\caption{The post-collision diagnostics for Fig.~\ref{fig:fig4} at the final time. The fractions $f_c$ and $f_r$ are defined in the text.}
\label{tab:collision_diagnostics}
\begin{ruledtabular}
\scriptsize
\begin{tabular}{lcccc p{0.30\columnwidth}}
case & $N$ & $\delta$ & $f_c$ & $f_r$ & outcome \\
Fig.~\ref{fig:fig4}(a) & 1 & 0 & 0.95 & 0.05 & coalescence-like \\
Fig.~\ref{fig:fig4}(b) & 5 & 0 & 0.61 & 0.32 & coalescence-like \\
Fig.~\ref{fig:fig4}(c) & 10 & 0 & 0.38 & 0.09 & coalescence-like \\
Fig.~\ref{fig:fig4}(d) & 1 & $\pi$ & $<10^{-3}$ & $<10^{-3}$ & quasi-elastic \\
Fig.~\ref{fig:fig4}(e) & 5 & $\pi$ & $<10^{-3}$ & $<10^{-3}$ & quasi-elastic \\
Fig.~\ref{fig:fig4}(f) & 10 & $\pi$ & $<10^{-3}$ & 0.01 & quasi-elastic
\end{tabular}
\end{ruledtabular}
\end{table}

In the in-phase set of Figs.~\ref{fig:fig4}(a--c), all three collisions leave a central localized remnant rather than a purely elastic two-packet output. The central remnant population provides a density-based indicator of coalescence-like dynamics, consistent with droplet-collision analyses that classify outcomes by the atom number staying near the collision point~\cite{Ferioli2019Collisions,Hu2022SymmetricCollisions}. The case ($N=1$) has the largest central retention and a visible breathing remnant, indicating transfer of the translational kinetic energy into inner vibrations of the emerging bound state. Such residual shape oscillations are a common signature of post-collision excitation in the bound state~\cite{Cikojevic2021Collisions,Hu2022SymmetricCollisions}. For the cases ($N=5$) and ($N=10$), smaller central fractions and outgoing density show that a part of the injected energy is carried away by outgoing packets and low-amplitude radiation. This behavior is consistent with the known results for the inelastic collisions between droplets and flat-top-solitons, where radiated norm and outgoing density fragments quantify the inelastic component~\cite{Astrakharchik2018Droplets,AlbaArroyo2022Weber,AlKhawaja2024FlatTop}. The central objects are therefore interpreted as coalesced excited remnants, rather than merged ground states.

Figs.~\ref{fig:fig4}(d--f) show the out-of-phase set at the same incidence speed. The imposed phase difference suppresses the constructive density overlap at impact, reduces the central retention below $10^{-3}$ in all three cases, and yields quasi-elastic separation with small radiation fractions. Such phase sensitivity is consistent with the results for collisions between quasi-1D droplets and flat-top solitons, where the out-of-phase impact suppresses the overlap whereas the in-phase impact promotes stronger deformation, radiation, and coalescence-like response~\cite{Otajonov2024Quasi1D,AlKhawaja2024FlatTop,Alotaibi2026PhaseControlled}. Auxiliary in-phase runs at $v_0=0.3$ for the intermediate and flat-top droplets shift the response to weakly inelastic collisions, which supports a gradual velocity-dependent crossover rather than a sharp critical velocity. Taken together, these results provide a manifestation of the crossover identified above. The soliton-like to flat-top transition modifies how the in-phase impact stores translational energy, whereas setting $\delta=\pi$ suppresses density overlap and drives the same states toward quasi-elastic separation.
\section{Conclusion}
We have formulated an effective 1D description of polar spin-1 QDs in which the LHY correction separates into density and spin fluctuation channels. The main physical result is that these attractive 1D fluctuation corrections can support self-bound droplets even when the spin-independent mean-field interaction is repulsive. The stationary branch evolves continuously from soliton-like finite-size states to flat-top droplets, accompanied by saturation of the peak density and chemical potential rather than by a sharp critical particle number. The BdG spectra and weak-perturbation dynamics support the stability of the computed branch in the explored regime. The quadratic-Zeeman quench further shows that the breathing response changes across the finite-size crossover and is sensitive to whether the density or spin fluctuation channel is varied. Head-on collisions reveal a complementary nonlinear response, where the relative phase controls whether the impact favors coalescence-like central retention or quasi-elastic separation. These results identify polar spin-1 droplets as a useful setting for separating the equilibrium and nonequilibrium roles of density and spin fluctuation channels in low-dimensional self-bound matter.

\appendix
\section{Polar Bogoliubov theory and quasi-1D reduction}
\label{app:lhy_derivation}
We start from a uniform spin-1 condensate with contact interactions and a quadratic Zeeman shift,
\begin{equation}
\begin{split}
\hat H=&\int d^3{\bf r}\Bigg\{
\sum_{m=-1}^{1}\hat\psi_m^\dagger\left(-\frac{\hbar^2\nabla^2}{2M}\right)\hat\psi_m
\\
&\qquad+\frac{c_0}{2}:\hat n^2:
+\frac{c_1}{2}:\hat{\bf F}^2:
+q(\hat n_{+1}+\hat n_{-1})
\Bigg\}.
\end{split}
\end{equation}
The linear Zeeman term is omitted because the longitudinal magnetization is fixed to the \(M_z=0\) sector, where it only adds an irrelevant constant. For \(c_0>0\), \(c_1>0\), and \(q>0\), the mean-field ground state is the polar spinor \(\bm{\zeta}_P=(0,1,0)^T\), with \(\mathcal E_{\rm MF}=(c_0/2)n_{\rm 3D}^2\) and \(\mu_{\rm MF}=c_0n_{\rm 3D}\). Expanding \(\hat\psi_0=\sqrt {n_{\rm 3D}}+\delta\hat\psi_0\) and \(\hat\psi_{\pm1}=\delta\hat\psi_{\pm1}\), and keeping terms up to quadratic order, gives one density branch and two degenerate transverse-spin branches,
\begin{equation}
\begin{aligned}
E_d(k)&=\sqrt{\varepsilon_k(\varepsilon_k+2c_0 n_{\rm 3D})},
\\
E_s(k)&=\sqrt{(\varepsilon_k+q)(\varepsilon_k+q+2c_1 n_{\rm 3D})},
\end{aligned}
\end{equation}
where \(\varepsilon_k=\hbar^2k^2/(2M)\). The twofold spin degeneracy is the origin of the factor of two in the spin-channel zero-point energy.

In the quasi-1D setting, transverse motion is frozen to the harmonic ground state with transverse frequency \(\omega_\perp\) and oscillator length \(a_\perp=\sqrt{\hbar/(M\omega_\perp)}\). Projecting onto this transverse mode gives the weak-coupling 1D interaction \(g_F^{1D}\simeq2\hbar\omega_\perp a_F\), valid for \(|a_F|/a_\perp\ll1\) away from a confinement-induced resonance~\cite{Olshanii1998}. The effective density and spin couplings are \(c_0^{1D}=(g_0^{1D}+2g_2^{1D})/3\) and \(c_1^{1D}=(g_2^{1D}-g_0^{1D})/3\).  The quasi-1D model is used under the self-consistency condition that the chemical potential, the local mean-field and LHY energy scales, and the low-lying collective-mode energies remain well below \(\hbar\omega_\perp\).

The corresponding zero-point energy is obtained after the standard Bogoliubov subtraction of the asymptotic single-particle terms. In the 1D weak-coupling limit, the spectra entering the zero-point integrals are the reduced 1D spectra at the local 1D density \(n_{\rm 1D}\),
\[
\begin{aligned}
E_d^{1D}(k)&=\sqrt{\varepsilon_k(\varepsilon_k+2c_0^{1D}n_{\rm 1D})},\\
E_s^{1D}(k)&=\sqrt{(\varepsilon_k+q)(\varepsilon_k+q+2c_1^{1D}n_{\rm 1D})} .
\end{aligned}
\]
This gives the channel-resolved zero-point integrals
\begin{equation}
\begin{aligned}
\mathcal E_{\rm LHY}^{1D,P}(n_{\rm 1D})=&
\int_0^\infty\frac{dk}{2\pi}\Big[E_d^{1D}(k)-(\varepsilon_k+c_0^{1D}n_{\rm 1D})\Big]
\\
&+\int_0^\infty\frac{dk}{\pi}\Big[E_s^{1D}(k)-(\varepsilon_k+q+c_1^{1D}n_{\rm 1D})\Big].
\end{aligned}
\end{equation}
Differentiation with respect to density gives the local LHY chemical potential
\begin{equation}
\begin{aligned}
\mu_{\rm LHY}^{1D,P}(n_{\rm 1D})=
&-\frac{1}{\pi}\frac{\sqrt{M}}{\hbar}(c_0^{1D})^{3/2}\sqrt{n_{\rm 1D}}
\\
&+\frac{1}{\pi}\frac{\sqrt{M}}{\hbar}(c_1^{1D})^{3/2}\sqrt{n_{\rm 1D}}\,
S_{1D}\!\left(\frac{q}{c_1^{1D}n_{\rm 1D}}\right).
\end{aligned}
\label{eq:lhy_mu_1d_polar}
\end{equation}
Equation~\eqref{eq:lhy_mu_1d_polar} separates the LHY term into density and spin fluctuation channels, with \(\gamma=q/(c_1^{1D}n_{\rm 1D})\). The auxiliary function \(S_{1D}(\gamma)\) is defined by
\[
S_{1D}(\gamma)=\frac{3}{2}I_{1D}(\gamma)-\gamma I_{1D}'(\gamma),
\]
with
\begin{equation}
\begin{aligned}
I_{1D}(\gamma)=\int_0^\infty dx\,\Big[&
\sqrt{(x^2/2+\gamma)(x^2/2+\gamma+2)}
\\
&-(x^2/2+\gamma)-1\Big].
\end{aligned}
\end{equation}
The density dependence in Eq.~\eqref{eq:lhy_mu_1d_polar} recovers the standard 1D \(n_{\rm 1D}^{3/2}\) scaling of the beyond-mean-field contribution.

 In the parameter range used in the numerical calculations, \(S_{1D}(\gamma)<0\), so the density- and spin-channel LHY terms are both attractive. Self-binding is therefore produced by the balance between the repulsive mean-field contribution and the total attractive 1D fluctuation correction. The reference choice \((\eta_0,\eta_1)=(1,1)\) is used to resolve the channel mechanism, while natural sodium-like parameters would make the spin-channel contribution much weaker.

 To give a representative physical scale, we consider a \(^{23}{\rm Na}\)-mass quasi-1D system with \(\omega_\perp=2\pi\times20\,{\rm kHz}\) and effective couplings \(c_0^{1D}/h=c_1^{1D}/h\simeq1.1\times10^{-4}\,{\rm Hz\,m}\simeq110\,{\rm Hz\,\mu m}\). This is an effective reference set with comparable density- and spin-channel couplings, not the unmodified natural sodium scattering point. The density-channel scale is close to the natural sodium order of magnitude, but a natural sodium-like limit has \(c_1^{1D}/c_0^{1D}\ll1\). Thus the spin-channel contribution is weak unless the effective spin channel is engineered. The reference scale gives \(n_1\simeq2.2\times10^5\,{\rm m}^{-1}\), \(\xi_1\simeq7.4\,\mu{\rm m}\), \(t_0\simeq2\times10^{-2}\,{\rm s}\), \(q/h\simeq24\,{\rm Hz}\) for \(\tau_q=1\), and \(N_{\rm phys}\simeq1.65N\) for \(\eta_0=1\). These axial and Zeeman scales are far below the transverse confinement scale.

 Three-body recombination is not included as a dissipative term in the conservative eGPE calculations. For the representative scale above, a conservative quasi-1D estimate gives loss times much longer than the simulated breathing and collision windows. Reduced density and 1D correlations further suppress three-body recombination in the low-dimensional regime~\cite{Lavoine2021,AstrakharchikGiorgini2006}.

\begingroup
\section{Particle-number dependence of the finite-droplet spectrum}
\label{app:bdgN_spectrum}
This appendix gives the supplemental \(N\)-dependent BdG spectrum used to support the stability and breathing discussion in Sec.~III~B and Sec.~III~C. The calculation uses the same stationary states, BdG linearization, local equation of state, and channel parameters \((\eta_0,\eta_1,\tau_q)=(1,1,1)\) as the main-text stability analysis.

\begin{figure}[t]
\centering
\includegraphics[width=0.86\linewidth]{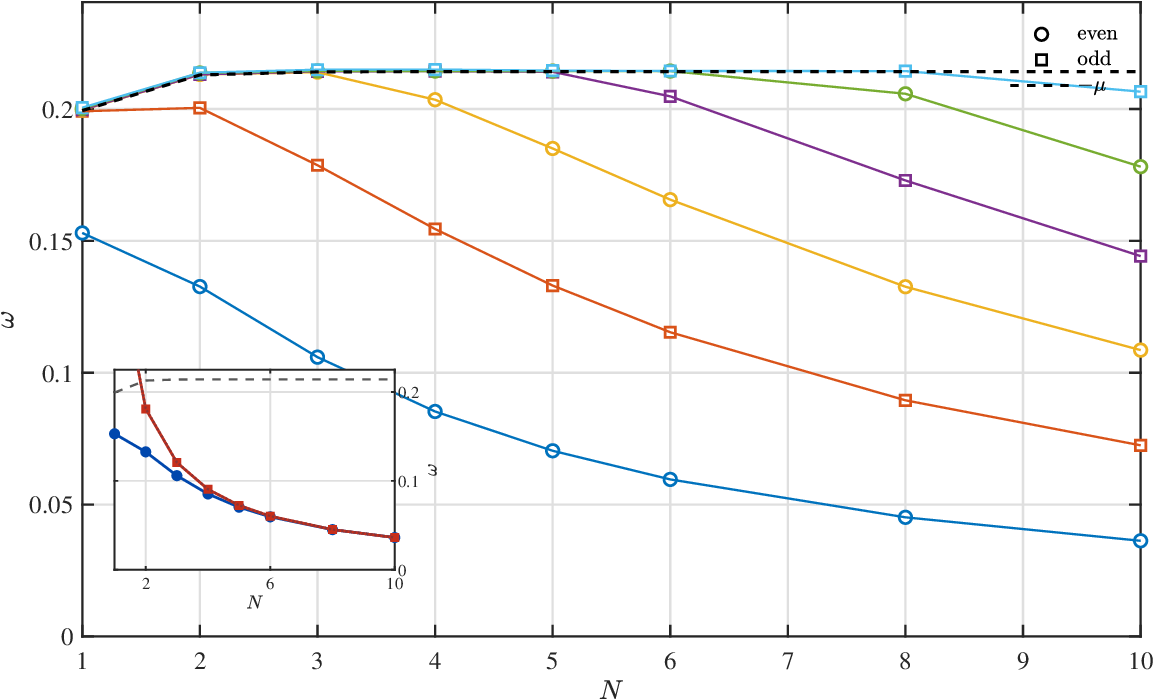}
\caption{Low-lying BdG eigenfrequencies of finite 1D polar spin-1 QDs as functions of particle number \(N\) at fixed \((\eta_0,\eta_1,\tau_q)=(1,1,1)\). Circles and squares denote even- and odd-parity density modes, respectively. The dashed line shows the particle-emission threshold \(-\mu\). The near-zero phase and translational modes are not included in the plotted finite-frequency modes. The inset compares the lowest even-parity density mode with the semi-analytic estimate \(A c_s/L_{\rm eff}\) and the reference value \(\pi c_s/L_{\rm eff}\).}
\label{fig:bdgN}
\end{figure}

For a finite self-bound 1D droplet, the axial wave vector is not a good quantum number. We therefore plot the discrete BdG eigenfrequencies as functions of \(N\) in Fig.~\ref{fig:bdgN}. The modes are classified by the parity of the density fluctuation. The lowest even-parity density branch decreases with \(N\), remains below the particle-emission threshold \(-\mu\), and has the strongest connection to the width observable \(\langle x^2\rangle\). We therefore use it as the breathing-relevant compressional branch. In one dimension, interfacial information is represented by the edge participation of finite-droplet density modes rather than by separate angular-momentum surface branches. The remaining branches are interpreted more cautiously as low-lying finite-droplet density modes. Modes close to \(-\mu\) are treated as threshold-adjacent. This presentation follows prior 1D-droplet work, where finite-droplet modes are compared with the particle-emission threshold~\cite{Tylutki2020,Du2023}.

The role of Fig.~\ref{fig:bdgN} in the manuscript is diagnostic. It does not by itself resolve the nonlinear modal content of the time evolution in Fig.~3. Instead, it supports the connection between the main-text stability spectra and the quench-induced breathing dynamics. The denser low-lying sector in the intermediate-size regime is consistent with stronger beating in the \(N=5\) dynamics, whereas the more separated low-frequency structure in the larger flat-top case is consistent with the more regular breathing pattern at \(N=10\). Thus the appendix spectrum connects to the breathing dynamics in Sec.~III~C through the spacing and parity of the low-lying density modes.

We further checked the lowest even-parity density mode with a semi-analytic liquid-segment estimate, as shown in the inset of Fig.~\ref{fig:bdgN}. This estimate uses the same local equation of state as the eGPE, namely the mean-field term together with the channel-resolved LHY term defined in Eqs.~\eqref{eq:lhy_mu_1d_polar} and \eqref{eq:egpe_dimless}, so that \(\mu_{\rm loc}(n)=3n-\eta_0\sqrt n+\eta_1\sqrt n\,S_{1D}(\tau_q/n)\). In the flat-top regime, the central density saturates. We therefore take the dimensionless bulk density from the large-\(N\) stationary states, \(n_{\rm b}\simeq0.31\). A long-wavelength density perturbation of a uniform 1D liquid segment has a sound speed determined by the slope of the local chemical potential,
\begin{equation}
c_s^2=n_{\rm b}\left.\frac{\partial\mu_{\rm loc}}{\partial n}\right|_{n_{\rm b}}.
\end{equation}
For \(n_{\rm b}\simeq0.31\) and \((\eta_0,\eta_1,\tau_q)=(1,1,1)\), this gives \(c_s\simeq0.38\). Since increasing \(N\) mainly enlarges the droplet rather than increasing the central density, the effective liquid length is estimated as
\begin{equation}
L_{\rm eff}\simeq \frac{N}{n_{\rm b}}.
\end{equation}
The lowest compressional sound-like mode of a finite liquid segment then has the scaling form
\begin{equation}
\omega_{\rm hyd}(N)=A\frac{c_s}{L_{\rm eff}}.
\end{equation}
Here \(A\) is a boundary-condition factor associated with reflection at the two droplet edges. We fix only this overall factor from the flat-top points \(N=8\) and \(N=10\),
\begin{equation}
A=\left\langle \omega_{\rm BdG}(N)\frac{L_{\rm eff}(N)}{c_s}\right\rangle_{N=8,10}
\simeq3.14\simeq\pi .
\end{equation}
The inset of Fig.~\ref{fig:bdgN} shows that this estimate tracks the lowest even-parity BdG frequency from \(N=5\) to \(N=10\). The agreement supports the conservative mode assignment in the flat-top regime and is consistent with prior 1D-droplet work on breathing-mode scaling and emission-threshold comparisons~\cite{Tylutki2020,Du2023,Orignac2024Breathing}.
\endgroup

\begin{acknowledgments}
This work was supported by the Natural Science Basic Research Program of Shaanxi Province (Grant No.~2026JC-JCQN-006), Zhiyuan Science Foundation of BIPT (Grant No.~2025207), the Beijing Institute of Petro-Chemical Technology urt Program (Grant Nos.~2025J00087, 2025J00088), the National Natural Science Foundation of China (Grant Nos.~12475004, 12175129, 12305029, 12334012, 12234012, and 52327808), the National Key R\&D Program of China (Grant Nos.~2024YFF0726700, 2021YFA1400900, and 2021YFA0718300), the Space Application System of China Manned Space Program, and the Elite Revitalizing Inner Mongolia Program (Grant No.~2025TGL05).
\end{acknowledgments}

\end{document}